\documentclass[conference]{IEEEtran}
\IEEEoverridecommandlockouts
\usepackage[utf8]{inputenc}
\usepackage{graphicx}
\usepackage{float} 
\usepackage{tikz}
\usepackage{xcolor}
\usepackage{caption}
\usepackage{subcaption}
\usepackage{hyperref}
\usepackage{placeins}
\usepackage{comment}
\usepackage{booktabs} 
\usepackage{multirow}
\usepackage{svg}

\usepackage{makecell}


\usepackage{amsfonts}
\usepackage{pifont}

\newcommand{\xmark}{\textcolor{red}{\ding{55}}}

\definecolor{lime}{HTML}{A6CE39}
\DeclareRobustCommand{\orcidicon}{
	\begin{tikzpicture}
		\draw[lime, fill=lime] (0,0)
		circle[radius=0.16]
		node[white]{{\fontfamily{qag}\selectfont \tiny \textbf {\.{I}D}}};
	\end{tikzpicture}
	\hspace{-2mm}
}
\foreach \x in {A, ..., Z}{%
	\expandafter\xdef\csname orcid\x\endcsname{\noexpand\href{https://orcid.org/\csname orcidauthor\x\endcsname}{\noexpand\orcidicon}}
}

\definecolor{dodgerblue}{RGB}{64, 67, 194}

\title{Role‑Based Agentic AI for Intent-driven \\ Network and Service Orchestration}

    \author{%
		\IEEEauthorblockN{
			\parbox{\linewidth}{\centering
				  Juan Parra-Ullauri* \hspace{-1.5mm}\orcidA{},
                Talha Ahmed Khan, Daniel McHugh, \\Shipra Kapoor, Alistair Duke, Alicia Hey, Andy Corston-Petrie
			}%
		}%
        \\\textit{ Research \& Commercialisation, BT Group, UK.}
        \thanks{*Corresponding author: Juan Parra-Ullauri (juan.m.parraullauri@bt.com)}
		\thanks{Disclaimer: The information presented in this paper is intended for research purposes only. The views, interpretations, and conclusions expressed herein are those of the authors and do not necessarily reflect the official position, policies, or opinions of BT Group or any of its affiliates.}
	}

\date{December}

\begin{document}

\maketitle

\begin{abstract}
Telecommunication networks are increasingly complex due to heterogeneous technologies, diverse service requirements, and growing demands for resource efficiency and business agility. Intent-Based Networking (IBN) and, more recently, agentic AI have emerged as promising paradigms to address this complexity through autonomous network management. 
However, existing approaches primarily focus on operational orchestration within Operations Support Systems (OSS) and lack an integrated framework that spans Business Support Systems (BSS) and OSS, limiting the realisation of true intent-to-business-to-network coordination.
This paper presents a role-based multi-agent architecture (MAS) for end-to-end intent orchestration that mirrors Communication Service Provider (CSP) organisational structures. The proposed framework applies principles of functional decomposition, explicit task ownership, privacy-preserving domain separation, and domain-specific expertise within a hierarchical four-layer agent system spanning customer engagement, strategic planning, service delivery, and infrastructure provisioning. Leadership agents coordinate planning activities, whilst specialised service and resource agents are dynamically instantiated according to intent requirements.
A proof-of-concept implementation demonstrates the feasibility of bridging the BSS–OSS divide through structured agent coordination, illustrating how agentic MAS can support accountable and scalable intent-driven service orchestration.
\end{abstract}


\section{Introduction}
\label{intro}

Telecommunications operators operate in increasingly heterogeneous, distributed environments with evolving service demands and pressure to reduce operational expenditure while enabling new revenue models. Traditional management approaches, based on static configurations and manual workflows, are poorly suited to the scale and variability of modern networks~\cite{maestro2024}. Intent-Based Networking (IBN) has emerged as a paradigm for translating high-level business objectives into service and resource configurations~\cite{kapoor2025genai}. However, existing IBN implementations largely focus on OSS-level automation and rarely incorporate Business Support Systems (BSS), limiting true end-to-end intent-to-business-to-network orchestration.

In parallel, advances in agentic AI introduce new models of distributed reasoning and goal-directed collaboration. Agentic systems powered by Large Language Models (LLMs) emphasise autonomous planning and coordinated execution across specialised agents, and have been applied to IBN (~\cite{maestro2024, brodimas2025agentic,wang2025confucius}). While such techniques have been explored for optimisation and closed-loop control, their integration across the full intent lifecycle remains immature. Current approaches often lack explicit governance structures, clearly defined role boundaries, privacy-preserving domain separation, and reproducible coordination patterns suitable for telecom-grade environments subject to stringent operational and regulatory requirements.

This article addresses these challenges through the design of a role-based, distributed multi-agent architecture for intent-driven orchestration. The framework mirrors the organisational structures of Communication Service Providers (CSPs), introducing functional decomposition, clear task ownership, privacy-preserving cross-domain collaboration, and domain-specific expertise within specialised agents. The architecture comprises three core elements: (i) a hierarchical multi-agent system (MAS) spanning customer engagement, strategic planning, service delivery, and infrastructure provisioning; (ii) a segmented knowledge plane enabling semantic interoperability while preserving domain data privacy; and (iii) integration with the BSS--OSS intent lifecycle. These components align with TM Forum\footnote{TM Forum, Open Digital Architecture (ODA)''} and ETSI Zero-touch Service Management (ZSM)\footnote{ETSI ISG ZSM, Zero-touch network and Service Management (ZSM); Reference Architecture,'' ETSI GS ZSM 002} principles for closed-loop automation, ensuring standards compliance and interoperability. Unlike existing frameworks, our approach embeds organisational role structures and governance constraints directly into agent coordination.

A proof-of-concept implementation demonstrates the feasibility of this approach using state-of-the-art agentic frameworks. ReAct-based reasoning loops~\cite{yao2023react} enable structured planning and action execution; the Model Context Protocol (MCP) provides controlled knowledge exchange and tool invocation across domains; and Agent-to-Agent (A2A) communication ensures secure, role-bound coordination between distributed agents. The implementation integrates with TMF Open APIs and ETSI-compliant orchestration components, illustrating standards-aligned interoperability across the BSS--OSS lifecycle.

The remainder of this article is organised as follows. Section~\ref{preliminaries} introduces the foundational concepts of IBN and MAS, and situates this work within the existing literature. Our proposed architecture is presented in Section~\ref{proposal}. Section~\ref{example} illustrates the framework’s operating principles through a concrete example, while Section~\ref{challenges} discusses research challenges and open issues. Finally, we conclude the article.

\section{Preliminaries and Related Work}
\label{preliminaries}
\subsection{Service Orchestration and Intent-Based Networking}

Service orchestration in telecommunications refers to the set of functions responsible for configuring and activating a \textit{network service} in response to customer requirements (intent). Intents, typically expressed as Service Level Agreements (SLAs), are decomposed into Customer Facing Services (CFS) and Resource Facing Services (RFS), and subsequently mapped to network resources to achieve service fulfilment. This process enables resource allocation, activation, and continuous assurance to maintain contractual performance alignment. Despite these structured workflows, fulfilment and assurance processes within many CSPs remain reliant on expert-driven interpretation, cross-domain reasoning, and bespoke configuration procedures, limiting agility and increasing operational overhead.

IBN introduces a declarative paradigm in which high-level goals are translated into policies and configurations that enable autonomous orchestration and compliance monitoring. IBN supports progression from script-based automation towards self-serving, self-fulfilling, and self-assuring network behaviour across the service lifecycle. However, achieving higher levels of autonomy requires predictive and adaptive optimisation across RAN, Core, transport, Edge, Cloud, and OSS/BSS domains. IBN provides translation and policy guardrails, but full autonomy depends on integration with intelligent capabilities for inference, learning, and distributed decision-making.

\subsection{AI and Multi-Agent Systems in Network Management}

 AI and multi-agent systems offer capabilities to IBN that could enable more autonomous, goal-driven network management, yet their integration across the full intent lifecycle remains limited. AI/ML techniques support perception of network behaviour, context-aware decision-making, and predictive optimisation across multiple domains, enabling partial alignment with high-level business intents \cite{8994961}. Prior work has demonstrated the potential of AI in IBN, including knowledge-graph reasoning for semantic validation of 5G Core configurations \cite{10946508} and reinforcement learning for distributed optimisation in dynamic wireless environments \cite{9372298}. Agentic AI extends these approaches by introducing autonomous agents capable of reasoning over intents, invoking tools, executing multi-step plans, and adapting actions in real time. While promising for near-zero-touch functions such as fault detection, root-cause analysis, performance prediction, and network-level resource orchestration, current solutions predominantly target isolated network-layer or OSS-layer tasks, leaving end-to-end business-to-network alignment underexplored.

\subsection{Related Work}

\begin{table*}[t]
\caption{Comparison of Agentic AI Approaches for Intent-Based Service Orchestration}
\label{tab:comparison}
\centering
\small
\begin{tabular}{lccccccc}
\toprule
\textbf{Work} & 
\textbf{IBN} & 
\makecell{\textbf{Multi-}\\\textbf{Agent}} & 
\makecell{\textbf{BSS--OSS}\\\textbf{Integration}} & 
\makecell{\textbf{Org.-Aligned}\\\textbf{Roles}} & 
\makecell{\textbf{Hierarchical}\\\textbf{Structure}} & 
\makecell{\textbf{Governance/Task}\\\textbf{Ownership}} & 
\makecell{\textbf{Lifecycle}\\\textbf{Coverage}} \\
\midrule
Brodimas \textit{et al.}~\cite{brodimas2025agentic} 
& \checkmark & \checkmark & \xmark & \xmark & Partial & \xmark & \makecell{Layered\\(OSS)} \\
Wang \textit{et al.}~\cite{wang2025confucius} 
& \checkmark & \checkmark & \xmark & \xmark & Partial & \xmark & \makecell{Layered\\(OSS)} \\
Guo \textit{et al.}~\cite{guo2025intent} 
& \checkmark & Partial & \xmark & \xmark & Partial & \xmark & \makecell{Layered\\(OSS)} \\
Chatzistefanidis \& Leone~\cite{maestro2024} 
& \checkmark & \checkmark & \xmark & \xmark & Partial & \xmark & \makecell{Layered\\(OSS)} \\
Sun \textit{et al.}~\cite{sun2026generative} 
& \checkmark & \checkmark & \xmark & \xmark & Partial & \xmark & \makecell{Layered\\(Edge)} \\
\midrule
\textbf{This paper} 
& \checkmark & \checkmark & \checkmark & \checkmark & \checkmark & \checkmark & \makecell{\textbf{E2E}\\\textbf{(BSS--OSS)}} \\
\bottomrule
\end{tabular}
\end{table*}

Closer to this work, recent advances in IBN and LLM-driven automation have stimulated research into agentic AI for network management. Brodimas \textit{et al.}~\cite{brodimas2025agentic} propose a distributed MAS that leverages LLM reasoning to translate high-level intents into coordinated infrastructure and service actions. Wang \textit{et al.}~\cite{wang2025confucius} introduce Confucius, a production-ready MAS framework that models network workflows as DAGs, integrates retrieval-augmented generation, and enforces validation for hyper-scale network operations. Guo \textit{et al.}~\cite{guo2025intent} design an autonomous networking architecture that abstracts user intents into resource allocation strategies using LLM-guided orchestration. Chatzistefanidis and Leone~\cite{maestro2024} develop Maestro, which applies LLM-driven collaborative automation for Open RAN optimisation and QoS-aware intent enforcement. Sun \textit{et al.}~\cite{sun2026generative} explore generative intent prediction with agentic AI to orchestrate edge service function chains dynamically.

Whilst these works show the feasibility of integrating LLMs and MAS within IBN, they mainly focus on OSS-layer orchestration. Integration of BSS within the intent lifecycle remains limited. Existing approaches generally organise agents by technical functionality rather than CSP organisational roles, and do not explicitly address governance, task ownership, or privacy-preserving domain separation. In contrast, this work proposes a hierarchical, role-based multi-agent architecture spanning customer engagement to service provisioning, enabling true end-to-end intent-to-business-to-network orchestration. Table~\ref{tab:comparison} summarises representative agentic IBN frameworks and highlights architectural differences in lifecycle coverage, organisational alignment, and governance modelling.

\section{Proposed Architecture}
\label{proposal}

This section describes the proposed role-based architecture, focusing on four key motivations: (i) functional decomposition through specialised roles, (ii) clear task ownership and accountability, (iii) privacy preservation across organisational boundaries, and (iv) leveraging domain-specific expertise. The architecture comprises three foundational components: a distributed MAS, a shared knowledge plane, and seamless integration with the OSS/BSS intent lifecycle stack. The objective is to establish a structured and reproducible pipeline for realising end-to-end intent-to-business-to-network orchestration.
\subsection{Role-Based Multi-Agent Framework}
\begin{figure*}
    \centering
    \includegraphics[width=\textwidth]{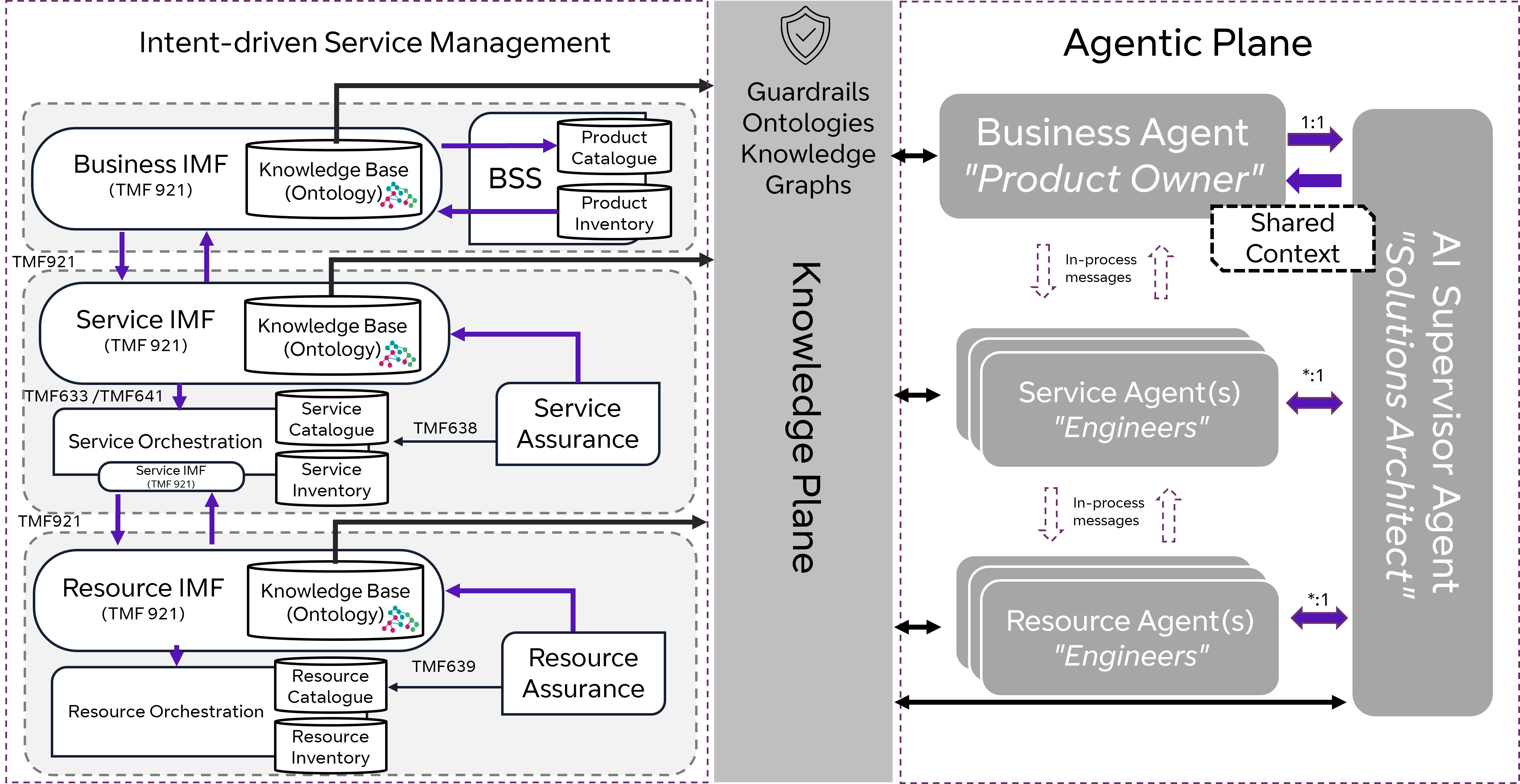}
    \caption{Role-based architectural framework for agentic intent-based service orchestration}
    \label{fig:architecture} 
\end{figure*}

Adopting an organisational perspective, the architecture (shown in Figure~\ref{fig:architecture}) implements a hierarchical MAS with clearly defined scopes and role assignments that mirror the operational structure of typical CSP delivery organisations. The proposed framework employs layered decomposition, wherein leadership agents coordinate strategic planning whilst specialised agents execute operational tasks at the service and resource layers. The agent hierarchy is structured as follows:

\subsubsection{Business Agent -- ``The Product Owner"}
The Business Agent serves as the initial point of contact in transforming customer requests (intents) into actionable specifications. Analogous to a product owner in agile methodologies, this agent operates at the business-customer interface, performing three primary functions: (i) customer engagement and requirements elicitation, (ii) intent refinement and validation, and (iii) representation of the CSP's service catalogue and business constraints. The agent's workflow begins with interactive requirement gathering, where ambiguous or incomplete customer requests are systematically refined through iterative dialogue. Once requirements are clarified and validated, the business agent formalises them into a product specification that serves as the contract for downstream processing across the orchestration stack.

Subsequently, the business agent initiates collaboration with the Supervisor Agent to develop and dimension the solution based on shared contextual knowledge. During this planning phase, feasibility assessments may reveal the need for requirement adjustments, either due to incomplete information or technical and business constraints. In such cases, the business agent re-engages with the customer to negotiate modifications to the original request, ensuring alignment between customer expectations and deliverable capabilities.

\subsubsection{Supervisor Agent -- ``The Solutions Architect"}
The Supervisor Agent receives the formalised intent and contextual information from the business agent. Considering business requirements, priorities, and organisational capabilities, it develops a comprehensive execution plan. The Solutions Architect handles three core tasks: (i) requirements interpretation and validation, (ii) end-to-end workflow decomposition, and (iii) definition of expected outcomes with quantifiable metrics for optimisation and lifecycle evaluation. A key function is agent orchestration and resource allocation, identifying participating agents, estimating capacity, and projecting engagement duration to calculate operational effort and costs. The Solutions Architect then translates these requirements into a structured execution plan, typically represented as a DAG, defining task dependencies, inter-agent relationships, and interaction patterns across the orchestration layers.

From an agentic AI perspective, the Supervisor Agent functions as the master orchestrator whilst additionally performing resource dimensioning for the MAS itself. This includes provisioning the infrastructure required to support agent collaboration, such as inter-agent communication channels, computational resources (e.g., GPU allocation per agent cohort), and network throughput requirements. The Supervisor Agent enforces organisational policy compliance at design time, embedding regulatory constraints, business rules, and operational guardrails directly into the execution plan. These policies are propagated across all layers of the stack, ensuring that subsequent agent actions remain within approved boundaries throughout the intent lifecycle.

\subsubsection{Service Agents}
Operating at the service layer, Service Agents constitute a pool of specialised agents, each responsible for specific service domains or functional capabilities. Unlike the singular Business and Supervisor Agents, multiple Service Agents may be instantiated and coordinated for any given intent, depending on the complexity and scope of the request. The Supervisor Agent dynamically selects and composes the appropriate Service Agents based on the requirements analysis and execution plan. Service Agent specialisations may include intent-management agents, security and authentication agents, billing and charging agents, and service assurance agents. Each specialised agent maintains deep expertise within its domain, including relevant service templates, configuration patterns, policy frameworks, and integration protocols.

The Service Agents perform domain-specific functions: (i) translating abstract service requirements into concrete service configurations within their specialisation, (ii) coordinating with peer Service Agents when service composition or cross-domain integration is required, (iii) generating service-level parameters and policies that guide resource-layer provisioning, and (iv) monitoring domain-specific KPIs against success metrics defined by the Supervisor Agent. Service Agents maintain awareness of available service building blocks, standard service templates, and composition rules specific to their domains. They abstract infrastructure complexity from the business layer whilst providing sufficient specificity for resource provisioning, translating business-oriented service descriptors into technology-agnostic service models that can be instantiated across heterogeneous infrastructure.

\subsubsection{Resource Agents}
At the infrastructure layer, Resource Agents form a diverse ecosystem of specialised agents, each responsible for specific technology domains, vendor platforms, or resource types. The architecture supports multiple concurrent Resource Agents, with the specific set engaged for any intent determined by the service requirements and infrastructure scope identified during planning. Resource Agent specialisations align with telecom infrastructure domains, including RAN agents for different radio technologies, core network agents, transport network agents for optical and IP/MPLS layers, virtualised infrastructure agents for NFV platforms, and cloud-native agents for containerised workloads. Each Resource Agent performs specialised operations: (i) translating service-level requirements into domain-specific resource configurations, (ii) executing provisioning operations through direct interaction with network elements and infrastructure APIs, (iii) monitoring resource-level telemetry and performance metrics, and (iv) reporting resource state and capability information to Service Agents and the Supervisor Agent.

Resource Agents maintain specialised knowledge bases containing domain-specific information such as equipment capabilities, vendor-specific configuration syntax, capacity constraints, and topology relationships. This distributed knowledge architecture enables the system to scale across heterogeneous multi-vendor environments without requiring centralised intelligence, with the number and type of Resource Agents activated varying significantly based on intent complexity.

\subsection{Knowledge Plane}

The knowledge plane serves as a foundational layer that enables structured, context-aware decision-making across the multi-agent architecture. It provides agents with semantically rich, domain-specific information, allowing them to understand their environment, reason about available resources, and execute tasks effectively. The knowledge plane leverages knowledge graphs to model entities, relationships, policies, and operational constraints, supporting inference, querying, and integration across service, resource, and business layers. This approach allows agents to pose targeted queries—for example, a Supervisor Agent may query whether sufficient RAN resources exist to fulfil a requested service without gaining direct access to low-level infrastructure details—preserving data privacy whilst providing actionable insights.

Crucially, the knowledge plane is architecturally segmented into multiple domain-specific planes that maintain ontological awareness of one another but do not share data universally. Each domain-specific plane controls access to its data, ensuring that only authorised agents within the appropriate scope can retrieve sensitive information. Agents in supervisory roles can frame requests to subordinate agents using semantically grounded terminology without requiring visibility into the underlying resource details. This segmentation prevents data leakage across organisational boundaries whilst enabling effective cross-domain coordination. Beyond data provision, the knowledge plane enforces guardrails and policies that define permissible agent behaviour, including regulatory compliance rules, business constraints, service-level policies, and operational boundaries embedded into the ontological framework. Agents rely on these guardrails to make autonomous decisions whilst remaining within approved limits, enabling safe, predictable, and policy-compliant execution across the BSS--OSS intent lifecycle.

Integration with the OSS--BSS stack is facilitated through the knowledge plane's semantic context. Business Agents leverage it to validate customer intents against catalogue definitions and organisational policies before formalisation, whilst Service and Resource Agents consult it during execution to ensure service configurations and infrastructure provisioning align with both intent and policy constraints. This tight coupling ensures that the knowledge plane not only supports informed decision-making and dynamic coordination amongst agents but also maintains the integrity, compliance, and consistency of end-to-end intent-to-network orchestration. The knowledge plane thus acts as the cognitive backbone of the agentic framework, enabling agents to operate autonomously yet cohesively, guiding interactions, decisions, and adaptations across the lifecycle whilst enforcing the organisational and regulatory boundaries defined in the architecture (see Figure~\ref{fig:architecture}).

\subsection{Intent Lifecycle and Integration Across OSS and BSS}  

Figure~\ref{fig:architecture} illustrates the integration of the proposed agentic framework with the OSS/BSS intent lifecycle stack, aligned with TMF-921 and extending the TMF Intent Ontology (TIO-ICM). The lifecycle begins in the BSS domain, where the Business Agent formalises a customer request into a structured, technology-agnostic intent, capturing both functional and non-functional requirements. This intent acts as a declarative contract specifying the desired outcome rather than the implementation method. Through the knowledge plane, it is validated against catalogue definitions, commercial policies, and organisational constraints before being propagated into the OSS domain. Within OSS, the intent is progressively refined across Business, Service, and Resource Intent Management Functions (IMFs), each maintaining its own lifecycle state whilst remaining semantically aligned through federated knowledge graphs. The Service IMF translates product-level objectives into service models and SLA parameters, whilst the Resource IMF converts service requirements into executable infrastructure configurations. Each layer independently compares intended versus operational state and generates structured intent reports, ensuring lifecycle separation whilst preserving a Single Source of Truth through ontological consistency rather than centralised data exposure.

Beyond traditional service assurance, the architecture establishes hierarchical intent assurance tightly coupled with the agentic execution model and the knowledge plane. Specialised agents at each layer continuously evaluate compliance between intended and observed states, enabling autonomous remediation within predefined policy guardrails, whilst the Supervisor Agent coordinates cross-layer adaptations when deviations have a broader impact. The knowledge plane provides the semantic context and boundary-aware data access required for agents to reason, negotiate, and reformulate intents without violating organisational separation. Intent thus becomes the governing abstraction that links distributed agents, domain-specific knowledge graphs, and OSS--BSS lifecycle management into a coherent, closed-loop, outcome-driven orchestration framework.

\section{Proof-of-Concept Implementation}
\label{example}
To demonstrate the feasibility of the proposed approach, we developed an implementation of the architecture encompassing the pipeline from customer intent expression and business-layer negotiation through to operational-layer execution (Figure~\ref{fig:system_design}). The intent is cascaded through the agent hierarchy, with each layer—Business, Supervisor, and Service—performing specialised transformations and orchestration tasks aligned with their designated roles. This section presents the experimental system design, implementation tooling, inter-agent information flows, and analysis of orchestration performance across the intent lifecycle.

\subsection{Experiment Design}

\begin{figure}
    \centering
    \includegraphics[width=\linewidth]{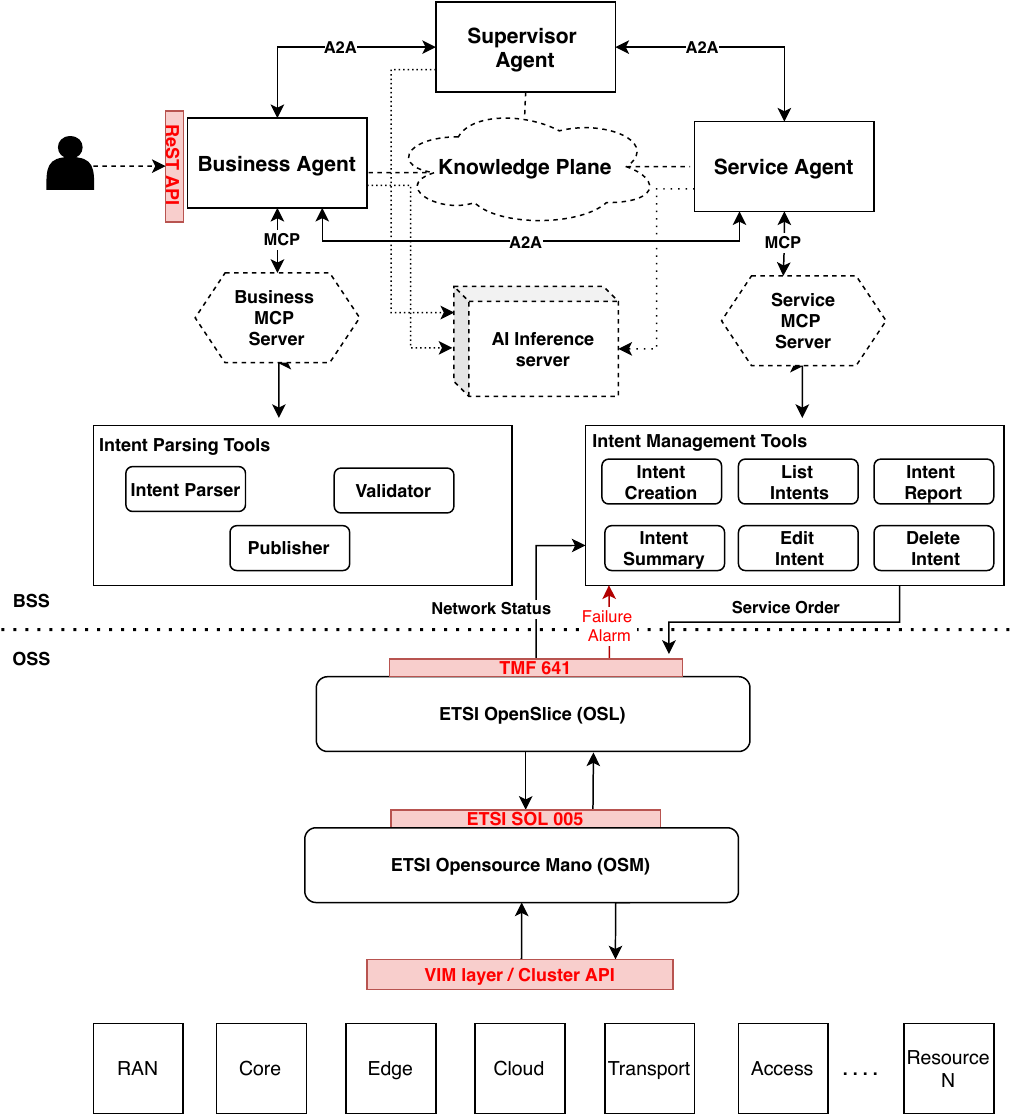}
    \caption{System architecture showing the integration of Business, Supervisor, and Service agents with MCP servers, knowledge plane, and BSS/OSS orchestration systems.}
    \label{fig:system_design}
\end{figure}

To validate the proposed architecture, we implemented three core agent instances: the Business Agent, the Supervisor Agent, and the Service Agent (Intent Handler). Figure~\ref{fig:system_design} illustrates the system architecture, showing the integration of these agents with the knowledge plane, MCP servers, and downstream orchestration systems including ETSI OpenSlice (OSL) and ETSI Open Source MANO (OSM). All agents are implemented using the ReAct framework within LangGraph, enabling structured planning and iterative tool invocation. LLM inference is provided by Ollama, hosting models in-house for deployment control and data privacy. Inter-agent communication is facilitated through A2A interfaces implemented using Starlette endpoints, whilst domain-specific capabilities are exposed via MCP servers that agents discover and invoke dynamically.

\subsubsection{Business Agent}

The Business Agent provides the customer-facing interface through a Gradio chatbot, connecting to the Business MCP Server using a MultiServer MCP client for dynamic tool discovery. During intent elicitation, the agent engages customers through iterative dialogue to refine requirements. The formalisation phase leverages two MCP tools: the ``Intent Parser'' tool transforms natural-language intent into structured RDF representation compliant with TMF Intent Ontology (TIO) v3.6.0 and Intent Common Model (TR290x), whilst the ``Validator'' tool performs syntactic and ontological validation using RDFLib. The generation-validation cycle repeats iteratively until the intent object passes validation, after which the Business Agent transmits the formalised intent to the Service Agent via A2A and confirms submission to the customer.

\subsubsection{Supervisor Agent}

The Supervisor Agent accesses the knowledge plane to retrieve organisational policies, current network status, and agent availability for feasibility assessment and workflow decomposition. Upon receiving a formalised intent from the Business Agent, it generates a structured execution plan identifying required specialised agents, task dependencies, and coordination patterns. The Supervisor communicates execution plans to participating agents via A2A interfaces, monitors progress throughout the intent lifecycle, and aggregates intentReports from subordinate agents to track fulfilment status.

\subsubsection{Service Agent (Intent Handler)}

The Service Agent operates as a continuously-running process exposing an A2A interface for asynchronous, event-driven orchestration. It dynamically discovers capabilities from the Service MCP Server to perform CRUD operations on intent objects and intentReports, reason over intent requirements using the Intent Knowledge Graph (IKG), and query intent state from the Intent MongoDB database. The Intent Handler maintains awareness of intent state throughout the lifecycle, proactively analysing feasibility, detecting conflicts through reasoning over the IKG, and taking corrective actions when needed. As shown in Figure~\ref{fig:system_design}, the Service Agent interfaces with ETSI OSL for service order management (TMF 641) and ETSI OSM for VNF lifecycle management (ETSI SOL 005), demonstrating standards-compliant integration across the orchestration stack.

\subsubsection{Knowledge Representation}

\begin{figure*}
    \centering
    \includegraphics[width=0.9\textwidth]{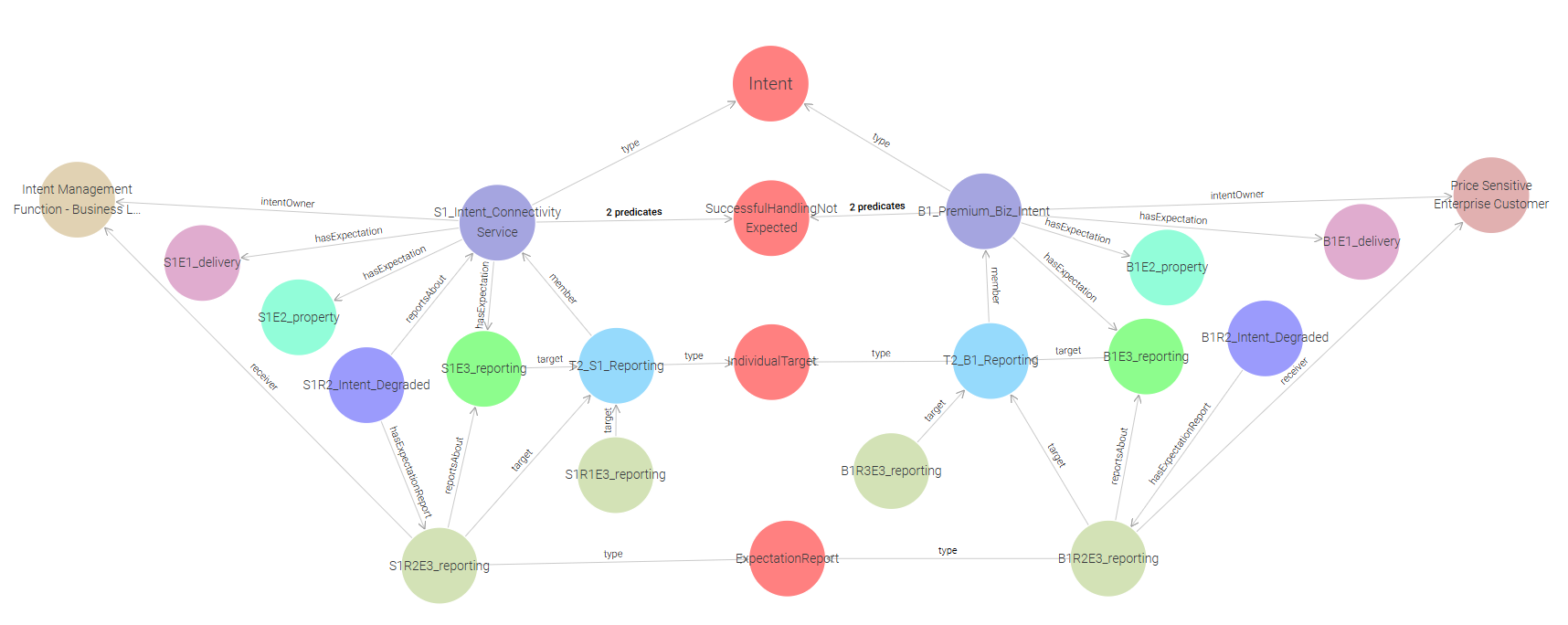}
    \caption{Intent Knowledge Graph ontology showing the semantic relationships between intent types (Business and Service), expectation models, reporting structures, and intent lifecycle states. The graph extends TMF Intent Common Model to support P2P connectivity scenarios.}
    \label{fig:ikg_ontology}
\end{figure*}

To instantiate the knowledge plane within our proof-of-concept implementation, we developed an Agentic Intent Management Framework (Agentic-IMF) that extends the baseline TMF Intent Management Framework (TMF-IMF) with MCP-exposed reasoning capabilities. The knowledge representation is structured around two core components: the IKG and an extended TMF Intent Ontology (TIO), both providing the semantic foundation for agent-based reasoning and autonomous assurance. Figure~\ref{fig:ikg_ontology} illustrates the ontological structure, which models relationships between intent entities, expectations, targets, and reporting mechanisms across the BSS--OSS layers. The ontology is rooted in the Intent Common Model (ICM) from TIO v3.6.0, extended to support point-to-point connectivity use-cases relevant to media production workflows. At the business layer, \texttt{B1\_Premium\_Biz\_Intent} represents customer-facing service requests with associated expectations, which decompose into service-layer specifications (\texttt{S1\_Intent\_Connectivity\_Service}) defining concrete service expectations and targets. The ontology captures bidirectional intent reporting flows: as service provisioning progresses, \texttt{ExpectationReport} instances document expectation fulfilment status, propagating upwards through the intent hierarchy to enable continuous state synchronisation across abstraction layers.

The Service Agent (Intent Handler) accesses the IKG through MCP-exposed tool interfaces, invoking SPARQL queries to retrieve intent state, validate expectation compliance, and update \texttt{intentReports} as orchestration activities complete. This enables ReAct agents to perform semantic reasoning over intent state, inferring actionable insights that would traditionally require rule-based logic or manual intervention. For instance, when an alarm indicates latency degradation, the Service Agent traverses the knowledge graph to identify the affected intent, determine which expectations are violated, assess severity based on business priorities encoded in the ontology, and autonomously trigger remediation actions such as resource reallocation. This approach provides autonomous assurance before requiring human intervention, enabling the multi-agent system to detect, diagnose, and resolve intent fulfilment issues through agentic inferencing rather than relying solely on predefined automation rules.

\subsection{Information Flow}
To illustrate the end-to-end orchestration workflow, we present a representative use case involving a media production company requesting dedicated connectivity with the intent: 

\colorbox{gray!20}{\parbox{0.9\linewidth}{\textit{``I need a dedicated high-bandwidth connection between our studio and our editing facility to transport uncompressed video. It must be highly reliable, low-latency, and offer guaranteed bandwidth. One-year term with monthly performance reports.''}}}
Fig.~\ref{fig:seq_diagram} illustrates the interaction workflow amongst the Customer, the Business Agent (Negotiator and Parsing), the MCP Business Server, the Supervisor Agent, the Service Agent (Intent Handler), and the MCP Service Server. The orchestration process unfolds through the following stages:

\begin{figure*}
    \centering
    \includegraphics[width=0.9\linewidth]{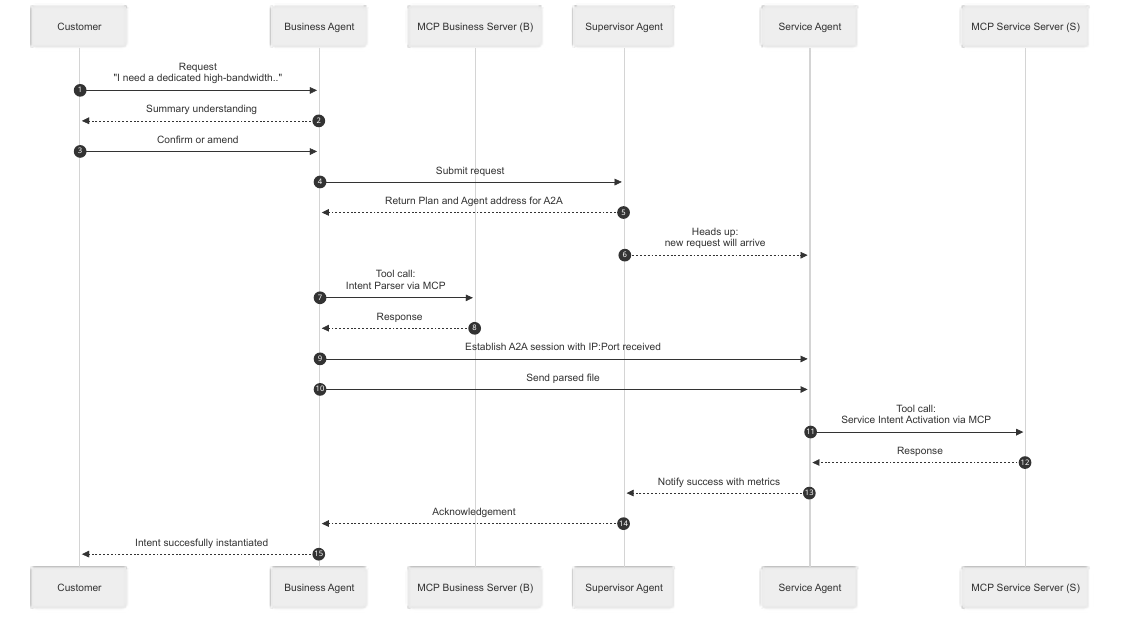}
    \caption{End-to-end sequence of customer request processing, from initial input through negotiation, to final service activation}
    \label{fig:seq_diagram}
\end{figure*}

\begin{itemize}
    \item Intent articulation and semantic confirmation:
    The Customer submits a natural-language request to the Business Agent (Step~1). The Business Agent generates a concise interpretation of the intent and returns it to the Customer for validation (Step~2). The Customer may confirm or amend this interpretation (Step~3), ensuring that subsequent processing operates on a stable and mutually understood intent description.

    \item Governance-guided admission and workflow preparation:
    Upon confirmation, the Business Agent forwards the request to the Supervisor Agent (Step~4). The Supervisor validates that the request is admissible and returns both a high-level execution plan and the communication endpoint required for A2A coordination (Step~5). In parallel, the Supervisor notifies the Intent Handler Agent of the impending request (Step~6), allowing downstream components to prepare for controlled orchestration.

    \item Business-level interpretation via MCP:
    The Business Agent invokes the MCP Business Server through a tool-call interface (Step~7) to obtain a structured business-domain interpretation. The Intent Parsing tools apply predefined business logic and return a machine-readable representation of the intent via the MCP Business Server(Step~8), forming the basis for service-domain translation.

    \item A2A session establishment and transmission of the parsed intent:
    Using the endpoint returned by the Supervisor, the Business Agent establishes an A2A session with the Intent Handler Agent (Step~9). The parsed intent file is then transmitted (Step~10), formally transferring responsibility for the request from the business domain to the service-orchestration domain. This step enforces separation of concerns and preserves privacy between organisational layers.

    \item Service-level activation through MCP:
    The Service Agent invokes the MCP Service Server (Step~11), which enables access to tools that perform the required service-orchestration procedures, including feasibility checks, resource allocation, and generation of service configurations that satisfy bandwidth, latency, and reliability constraints. The MCP Service Server returns its activation response to the Service Agent (Step~12).

    \item Completion reporting and hierarchical acknowledgement:
    After successful activation, the Service Agent reports completion to the Supervisor Agent and includes the relevant operational metrics (Step~13). The Supervisor then issues an acknowledgement to the Business Agent (Step~14), confirming that the request has been fulfilled in accordance with governance and operational rules.

    \item Customer notification:
    Finally, the Business Agent communicates to the Customer that the intent has been successfully instantiated (Step~15), concluding the end-to-end orchestration workflow and transitioning the service into its operational lifecycle.
\end{itemize}

\subsection{Results}

\begin{table*}[ht]
\centering
\caption{Agentic AI System Results}
\label{tab:agentic-results}

\begin{tabular}{cccccccccc}
\toprule
Agent &
\makecell{Inter-agent\\Messages} &
\makecell{Intra-agent\\Messages} &
\makecell{Total Tool\\Calls} &
\makecell{Total\\Tokens} &
\makecell{Avg\\Throughput} &
\makecell{Avg\\Latency} &
\makecell{Reasoning\\Steps} &
\makecell{Time to\\Complete} & \makecell{Successful}   \\
\midrule
\multicolumn{10}{c}{\textbf{gpt-oss:20b }} \\
Business Agent & 4 & 18 & 3 & 19,213 & 939.01 tok/s & 8.56 s & 6 & 44.24 s & Y\\
Service Agent & 6 & 216 & 16 & 145,354 & 1,218.90 tok/s & 7.58 s & 27 & 206.20 s & Y  \\
Supervisor Agent & 4 & 46 & 3 & 5,648 & 709.10 tok/s & 1.29 s & 8 & 2.12 s & Y \\
\midrule
\multicolumn{10}{c}{\textbf{mistral-nemo}} \\
Business Agent & 0& 30 & 0 & 5,989 & 601.79 tok/s & 2.28 s & 5 & 2.31 s & N \\
Service Agent& 6 & 309 & 18 & 95,728 & 1,135.05 tok/s & 4.64 s & 34 & 165.48 s & Y   \\
Supervisor Agent & 4 & 14 & 2 & 2,622 & 315.67 tok/s & 2.85 s & 4 & 3.35 s & Y \\
\midrule
\multicolumn{10}{c}{\textbf{qwen3:4b}} \\
Business Agent & 0 & 6 & 0 & 4,026 & 299.59 tok/s & 9.41 s & 2 & 9.46 s & N \\
Service Agent & 6 &291 & 18 & 129,798 & 356.65 tok/s & 16.80 s & 31 & 522.25 s & Y  \\
Supervisor Agent  & 0& 34 & 2 & 16,965 & 401.37 tok/s & 8.34 s & 6 & 16.82 s & N\\
\bottomrule
\end{tabular}

\end{table*}

To evaluate both individual agent capability and end-to-end integration, we conducted two complementary assessments. First, each agent was tested in isolation using prompt injection to emulate inter-agent messages, determining whether the agent could successfully complete its designated responsibilities when provided with correctly formatted inputs (marked as `Successful' in Table~\ref{tab:agentic-results}). Second, agents were deployed in the full orchestration pipeline to assess autonomous coordination via A2A interfaces (reflected in `Inter-agent Messages'). This dual methodology distinguishes between agent-level competence and system-level integration capability.

Table~\ref{tab:agentic-results} summarises the performance of three evaluated models (GPT-OSS, Mistral-Nemo, Qwen3:4B) across the Business, Service, and Supervisor Agents. The experiment replicated identical workflow configurations, prompting strategies, and system architectures for each model to isolate the impact of model capability on orchestration performance.

\textbf{Business Agent.} GPT-OSS successfully completed both isolation testing (marked Y) and pipeline integration (4 inter-agent messages, 3 tool calls), generating well-structured intent interpretations with moderate latency (8.56s). Both Mistral-Nemo and Qwen3:4B failed isolation testing (marked N), demonstrated by zero tool calls and zero inter-agent messages, preventing pipeline initiation entirely. Mistral-Nemo's failure is particularly notable given its competitive latency (2.28s), demonstrating that processing speed cannot compensate for functional inadequacy in critical parsing and MCP tool invocation tasks. Qwen3:4B exhibited even more severe limitations, combining functional failure with elevated latency (9.41s).

\textbf{Supervisor Agent.} GPT-OSS and Mistral-Nemo both succeeded in isolation (marked Y) whilst Qwen3:4B failed (marked N). Integration metrics show GPT-OSS and Mistral-Nemo both achieved 4 inter-agent messages, confirming successful governance coordination when properly invoked. Qwen3:4B recorded 0 inter-agent messages due to cascading failure from the Business Agent. Performance overhead remained minimal across successful executions (GPT-OSS: 2.12s, Mistral-Nemo: 3.35s), validating the lightweight nature of supervisory governance checks.

\textbf{Service Agent.} All three models succeeded in isolation testing (all marked Y), indicating fundamental capability to perform service-layer orchestration when provided with properly formatted inputs. GPT-OSS demonstrated reliable execution with 6 inter-agent messages, 16 tool calls, and 206.20s completion time, though elevated token generation (145,354 tokens) reflects verbose reasoning. Mistral-Nemo achieved optimal performance: 6 inter-agent messages, 18 tool calls, 165.48s completion time, and highest throughput (1,135.05 tok/s). Qwen3:4B, whilst functionally successful in isolation, exhibited severe performance degradation: 522.25s completion time (2.5× slower than GPT-OSS) driven by reduced throughput (356.65 tok/s), despite intermediate token volumes (129,798).

\textbf{Integration vs. Isolation Performance.} The divergence between isolation success and integration metrics reveals critical architectural insights. Mistral-Nemo's Service and Supervisor Agents succeeded in isolation but were never engaged in the integrated pipeline due to upstream Business Agent failure (0 inter-agent messages). This demonstrates that in hierarchical MAS, bottlenecks at initial intake layers can negate the capabilities of otherwise functional downstream components

Overall, the results indicate that larger or better-optimised models such as GPT-OSS are essential for the Business Agent layer, where semantic parsing, tool invocation, and pipeline initiation require robust language understanding. Smaller models may be viable for downstream components when invoked through controlled interfaces, provided their performance meets operational latency requirements. However, the inability of Mistral-Nemo and Qwen3:4B to reliably initiate the orchestration pipeline highlights the risks of deploying smaller models at customer-facing intake points in production MAS architectures.

\section{Open Challenges}
\label{challenges}
Whilst the experimental validation demonstrates technical feasibility, several ecosystem-level constraints remain barriers to production deployment of telecom-grade agentic automation. This section examines five critical challenge.

\subsection{Non-Determinism}
LLM-based agentic systems exhibit inherent non-deterministic behaviour, producing variable reasoning paths, tool invocations, and outputs even for identical inputs. This complicates reproducibility, debugging, and outcome verification in production workflows, which are critical requirements for telecom operations where service assurance demands predictable behaviour. Mitigation strategies include prompt engineering with structured output schemas, idempotency guarantees in tool design, and outcome-focused validation frameworks. Operators must shift from expecting deterministic execution paths to validating final outcomes against success criteria, supported by comprehensive logging, reasoning trace analysis, and model-specific operational expertise.

\subsection{Interoperability}
Telecom networks span multiple administrative domains, vendors, and standards bodies, producing heterogeneous data formats, inconsistent terminology, and substantial unstructured information. Effective agent reasoning requires semantic alignment across these information sources through ontology engineering and knowledge graph curation, processes that are resource-intensive and demand continuous maintenance as network technologies evolve. Whilst generative AI can accelerate ontology development and support explainable reasoning through natural language interfaces, generated knowledge representations remain error-prone and require rigorous human validation to ensure operational accuracy and trustworthiness.

\subsection{Validation and Verification}
Validating agentic outputs presents fundamental challenges due to non-deterministic reasoning and the existence of multiple valid solution paths for complex orchestration tasks. Established techniques include benchmark datasets, human-in-the-loop evaluation, input sanitisation, and self-evaluation methods such as chain-of-thought prompting and LLM-as-a-judge frameworks. However, these approaches face inherent limitations: post-hoc explanations frequently misrepresent actual reasoning processes, and increasing model complexity reduces interpretability. Our isolation testing methodology demonstrates a pragmatic approach: validating agents against defined success criteria rather than attempting to reproduce exact execution traces.

\subsection{Security}
AI agents introduce security vulnerabilities including prompt injection attacks, training data poisoning, jailbreaking attempts, hallucination-induced misconfigurations, and unauthorised access to privileged operations. Multi-agent systems amplify these risks through inter-agent message passing, shared knowledge planes, and cascading trust relationships. For telecom network operations, security-critical domains such as authentication systems, routing control, and customer data access must be strictly isolated from agentic automation. High-risk actions including network reconfigurations and policy modifications should mandate human-in-the-loop approval to prevent autonomous introduction of vulnerabilities.

\subsection{Costs and Energy Efficiency}
Deploying agentic AI in production telecom networks incurs substantial capital and operational expenditure. Requirements include specialised GPU infrastructure, redundant compute capacity for high-availability orchestration, and continuous model fine-tuning. Our experimental results illustrate these trade-offs: GPT-OSS achieved full pipeline success but generated 145,354 tokens over 206.20s, whilst Mistral-Nemo demonstrated 20\% faster completion with comparable reliability. Deployment decisions must balance model capability requirements against infrastructure costs, energy consumption, and sustainability targets. Efficient architectural design is essential for economically and environmentally sustainable production deployment.

\section{Conclusion}
\label{conclusion}
This paper presents a role-based, distributed multi-agent architecture for intent-driven orchestration in telecommunications networks, addressing limitations of existing IBN approaches that focus narrowly on operational automation without structured BSS--OSS integration. The proposed framework introduces hierarchical agent roles mirroring CSP structures, embedding clear task ownership, privacy-preserving cross-domain collaboration, and domain-specific expertise. The architecture comprises three foundational components: a hierarchical MAS spanning customer engagement through infrastructure provisioning, a segmented knowledge plane enabling semantic interoperability whilst preserving data privacy, and standards-compliant integration with TMF APIs and ETSI orchestration frameworks.

Our proof-of-concept implementation validates end-to-end orchestration from customer intent elicitation through service activation. Experimental results reveal that larger models such as GPT-OSS are essential for customer-facing Business Agents requiring robust semantic parsing and tool invocation, whilst smaller models may suffice for downstream coordination roles. Critically, failures at initial intake layers prevent otherwise capable downstream agents from being engaged, underscoring the importance of model selection at critical handoff points. Beyond architectural contributions, this work examines ecosystem-level challenges including non-deterministic behaviour, interoperability constraints, security vulnerabilities, and deployment costs. Our architecture enables CSPs to incrementally adopt AI automation without wholesale replacement of existing BSS--OSS infrastructure, reducing deployment risk whilst progressively realizing operational efficiency gains.

Future work should address comprehensive evaluation across diverse intent scenarios, advanced reasoning capabilities for complex cross-domain orchestration, formal verification methods for agent-generated configurations, and integration with closed-loop assurance mechanisms. The transition to agentic orchestration represents a paradigm shift for telecommunications operators, promising significant gains in automation and flexibility whilst demanding careful navigation of technical, organisational, and regulatory constraints.
\bibliographystyle{IEEEtran}
\bibliography{bibliography} 

\section*{Author Information}
\begin{IEEEbiographynophoto}{Juan Parra-Ullauri} [M] is an Electronics and Telecom Engineer with a PhD in Computer Science. He works as Research Manager in Intelligent Service Orchestration CoE at BT, and holds an Honorary Research position at the University of Bristol. Juan has led major UK and EU-funded research projects and received the National AI Award for High Tech \& Telecom (UK, 2024) as part of the REASON project. His expertise spans AI, Distributed Systems, and Data Engineering, with applications in Cloud, Telecom, and IoT networks.
\end{IEEEbiographynophoto}
\vspace{-30pt}
\begin{IEEEbiographynophoto}{Talha Ahmed Khan} is a Research Manager in Intelligent Service Orchestration at BT, with over three years of post-PhD experience in advanced research and development across academia and industry.
\end{IEEEbiographynophoto}

\vspace{-30pt}
\begin{IEEEbiographynophoto}{Daniel McHugh} is a Research Specialist and Thread Lead for Intent Based Networking within BT’s Intelligent Service Orchestration Centre of Excellence. He received his Bachelor’s degree in Digital Technology Solutions, with a specialism in Cyber Security, from the University of Exeter in 2024. His current research is focussed on intent-driven orchestration, advancing BT’s progression towards an intelligent autonomous network. 
\end{IEEEbiographynophoto}

\vspace{-30pt}
\begin{IEEEbiographynophoto}{Shipra Kapoor} is a Research Specialist and Thread Lead for Autonomous Orchestration in the Research \& Commercialisation group at BT, where she is leading high-impact research under the Intelligent Service Orchestration Centre of Excellence. 
\end{IEEEbiographynophoto}

\vspace{-30pt}
\begin{IEEEbiographynophoto}{Alicia Hey} is a Master's student in Digital Technology Solutions, specialising in Software Engineering. She has 4 years' experience at BT, and her current work draws focus to validation and verification of autonomous agentic systems.
\end{IEEEbiographynophoto}

\vspace{-30pt}
\begin{IEEEbiographynophoto}
{Alistair Duke} is a Research Manager in the Intelligent Service Orchestration Research Centre of Excellence within the Research \& Commercialisation group at BT. He has over 30 years experience in industrial research within the telecommunications sector including the fields of Internet of Things, Smart Cities, Semantic Web, Knowledge Management and Telepresence. 

\end{IEEEbiographynophoto}

 \vspace{-30pt}
\begin{IEEEbiographynophoto}{Andy Corston-Petrie} has more than 30 years of experience in the telecommunications industry.  A Physics graduate from the University of St Andrews, Andy leads the Intelligent Service Orchestration Research Centre of Excellence within the Research \& Commercialisation group at BT and has held a variety of positions at BT, including software development, system testing and programme delivery, and spent 10 years as the lead service designer of BT's IPX solutions.
\end{IEEEbiographynophoto}
    
	\vfill
\end{document}